\documentstyle[11pt,newpasp,twoside, epsf]{article}
\def\edcomment#1{\iffalse\marginpar{\raggedright\sl#1\/}\else\relax\fi}
\reversemarginpar

\begin{document}
\newcommand{\nhi}{\mbox{$N_{\rm HI}$}}
\newcommand{\hi}{H~{\sc i}}
\newcommand{\ahiss}{AH{\sc i}SS}
\newcommand{\himf}{H{\sc i}MF}
\newcommand{\mhi}{\mbox{$M_{\rm HI}$}}
\newcommand{\msol}{\mbox{${\rm M}_\odot$}}
\newcommand{\hmpc}{\mbox{$h_{100}^{-1}\, \rm Mpc$}}
\newcommand{\ihmpc}{\mbox{$h_{100}\, \rm Mpc^{-1}$}}
\newcommand{\kms}{\mbox{$\rm km\, s^{-1}$}}
\newcommand{\icmsq}{\mbox{$\rm cm^{-2}$}}
\newcommand{\rg}{\mbox{$r_{\rm g}$}}                       
\newcommand{\fnhi}{\mbox{$f(N_{\rm HI})$}}
\newcommand{\fn}{\mbox{$f(N)$}} 
\newcommand{\dndz}{\mbox{$dN/dz$}}

\title{Constraints on the Space Density of Extragalactic HVCs}
 \author{Martin A. Zwaan}
\affil{Kapteyn Astronomical Institute,
PO Box 800,
9700 AV Groningen,
The Netherlands}

\begin{abstract}
 High Velocity Clouds (HVCs) have recently attracted renewed attention
as being long lived,
massive dark matter dominated clouds of primordial composition
distributed throughout the
Local Group.  In this picture the HVCs would contain a few $\times
10^7~M_\odot$ of \hi\ and would be at distances of a few hundred kpc to
1.5~Mpc from the Local Group barycenter.  If this extragalactic
interpretation of HVCs is true, similar clouds are expected in other
galaxy groups and around galaxies.  We discuss the limits blind \hi\
surveys and QSO absorption line studies put on this proposed population
of clouds. 

\end{abstract}

\section{Introduction}
 The nature of high velocity clouds (HVCs) still remains a mystery. 
Several explanations of their origin have been set forth by Wakker \&
van Woerden (1997) in their review of forty years of HVC research. 
They conclude that no single origin can explain all observed HVCs and
subdivide all known HVCs into different classes.  Well established is the
explanation for the Magellanic Stream, the largest \hi\ structure
observed in the sky, except for the gaseous disk of the Milky Way
itself.  This HVC is likely the result of tidal interaction between the
Milky Way and the Magellanic Clouds, where LMC gas is stripped away and
spread along an orbit around the Milky Way.  However, this HVC is not
typical for the large number of clouds presented by Wakker \& van
Woerden (1991), the standard catalog of HVCs.  Most clouds require
alternative explanations. 

The main uncertainty in understanding the nature of the HVCs is that
their distances are difficult to determine.  The absence of starlight in
HVCs prevents the use of spectroscopic parallax methods commonly-used to
determine distances to stars in the nearest galaxies.  
Some distance brackets have been
specified with the absorption line method where spectra of stars with
known distances are checked for the presence of (e.g.) Ca, Fe, and Mg
absorption features (Van Woerden et al.  1999).  Recently, Balmer
recombination line emission driven by reprocessing of the UV ionizing
radiation originating from hot, young stars in the Galactic disk has
been used to specify distances to a few clouds (Bland-Hawthorn et al. 
1998, Bland-Hawthorn \& Maloney 1999).  These measurements require an
accurate model for the Galactic ionizing field, and work best for clouds
with large perpendicular distances from the Galactic disk.  Braun \&
Burton (1999) present yet another distance estimator.  For a cool HVC
core they use the measured \hi\ column density and the angular size to
calculate an \hi\ volume density which depends on the distance to the
cloud. Assumptions about the thermodynamics of the \hi\ gas yields a
likely density and hence a distance. 

Oort (1966) was the first to consider HVCs in an extragalactic context. 
His distance estimation assumed that the
clouds are self-gravitating, stable entities.  By comparing their solid
angle on the sky, brightness temperature, and velocity dispersion
Hulsbosch (1975) found that the typical distance would be approximately
10~Mpc, which places the clouds outside the Local Group (LG) and implies that
their neutral hydrogen mass would be $\sim 10^9\msol$, comparable to
that of normal spiral galaxies.  At these distances, the clouds should
be participating in the Hubble expansion, rather than approaching the
Milky Way.

Blitz et al.  (1999) and Braun \& Burton (1999) have recently revived the
idea that HVCs are extragalactic.  These authors show that the clouds'
distribution on the sky, as well as their kinematics as an ensemble are
consistent with a model in which HVCs are distributed throughout the
LG, at typical distances of a few hundred kpc (Braun \& Burton)
to 1~Mpc (Blitz et al.).  In order to make the total mass of the HVCs
consistent with the assumption of self-gravity, the authors suggest
that, like galaxies, HVCs are dominated by dark matter.  This dark
matter provides the binding potential that keeps the clouds from falling
apart; the only directly observable constituent, \hi, is only 10\% of
the total mass.  In this scenario, the HVCs are either remnants from the
formation of the LG or representatives from an intergalactic
population of dark matter dominated mini-halos in which hydrogen has
collected and remained stable on cosmological time scales. 

A theoretical basis for the LG explanation is provided by hierarchical
clustering scenarios that explain the formation of galaxies by many
generations of mergers of smaller masses.  Large numbers of
proto-galactic gas clouds might survive to the present day if the
mergers are not fully efficient.  However, only about ten percent of the
predicted number of halos have been identified as dwarf galaxies (Klypin
et al.  1999, Moore et al.  1999).  The association of HVCs with this
missing population makes an appealing picture for the LG explanation. 

Clearly, if the HVC phenomenon is a common feature of galaxy formation
and evolution, then extragalactic surveys of the halos and group
environments of nearby galaxies should show evidence for this
population.  Also the incidence of QSO absorption line systems should be
in agreement with this idea.

\section{The \hi\ Mass Function of HVCs in the Local Group}
 Determining the space density of gas-rich galaxies and possible
intergalactic gas clouds as a function of \hi\ mass has been, and
continues to be, one of the main objectives of extragalactic 21cm
surveys.  At this meeting the results from several \hi\ surveys with
varying sensitivity and survey volume have been presented, and it is
clear that at this moment there is still no consensus on the shape and
normalization of the \hi\ mass function (\himf, see contributions by
Henning, Schneider, Staveley-Smith, Verheijen, Webster).  Especially at
low \hi\ masses ($\mhi<10^{7.5}~h_{65}^{-2}~\msol$), there is
considerable uncertainty due to the small number of detections. 
Schneider, Spitzak \& Rosenberg (1998) have found evidence for a steep
upturn in the tail of the \himf.  Although this steep tail has a
tantalizing similarity to the signature of massive HVCs, it appears that
at least one of the two \hi\ signals responsible for the rise comes from
a normal galaxy, and the other is too close to a bright star to exclude
faint optical emission (Spitzak \& Schneider 1999). 
                         
In order to develop a reference frame for the number density of HVCs in
the LG, we determine the LG \himf\ from \hi\ measurements of all known
members as compiled by Mateo (1998).  The top panel of
Figure~1 shows the resulting \himf, constructed from 22 LG
galaxies in which \hi\ has been detected.  Also incorporated in this
mass function are the recent \hi\ detections in dwarf spheroidals (Blitz
\& Robishaw 2000).  A simple correction for incompleteness due to
obscuration by dust in the Zone of Avoidance has been made (see Zwaan \&
Briggs 2000).  Also shown is the field \himf\ determined by Zwaan et al.
(1997), scaled vertically so as to fit the points around the knee in the
\himf\, where the curve has been measured accurately.  Note that the
\himf\ of optically selected galaxies in the LG is remarkably flat, with
$\alpha\approx -1.0$.  Dwarf galaxies with $\mhi<10^8\msol$ contribute
only $\sim 2.5\%$ to the total \hi\ budget of the LG. 

The second panel of Figure~1 shows the same curve, but
overlaid is now the \himf\ of HVCs if they were self-gravitating clouds
distributed throughout the LG, as proposed by Blitz et al.  (1999).  We
use the HVC parameters from the compilation of HVCs of Wakker \& van
Woerden (1991).  For each cloud, the distance at which it is
gravitationally stable is calculated, assuming that the ratio of
baryonic to total mass if $f$.  We choose to let $f$ vary from $0.0125$
to $0.2$ and calculate the mass functions.  The virial distance of a
cloud scales in direct proportion to $f$, and hence the mass with $f^2$. 
It is obvious from Figure~1 that high values of $f$ are in
variance with the the observed field \himf: 21cm surveys should have
detected in the order of 10 dark \hi\ clouds for every normal galaxy
with the same \hi\ mass in the range $10^{7.5}$ to $10^{9}~\msol$.  None
of the \hi\ surveys to date has found evidence for such a large
population of \hi\ clouds without optical identification. 

The space density of HVCs in the LG can only be brought into agreement
with the observed field \himf\ if the median value of $f$ is lowered to
$\sim~0.02$, a value much lower than what is normally observed in
galaxies.  The median distance of such clouds must be smaller than $\sim
200$~kpc.  At these distances the clouds do not fit logically in a
model in which they are distributed throughout the LG.

\begin{figure} \plotone{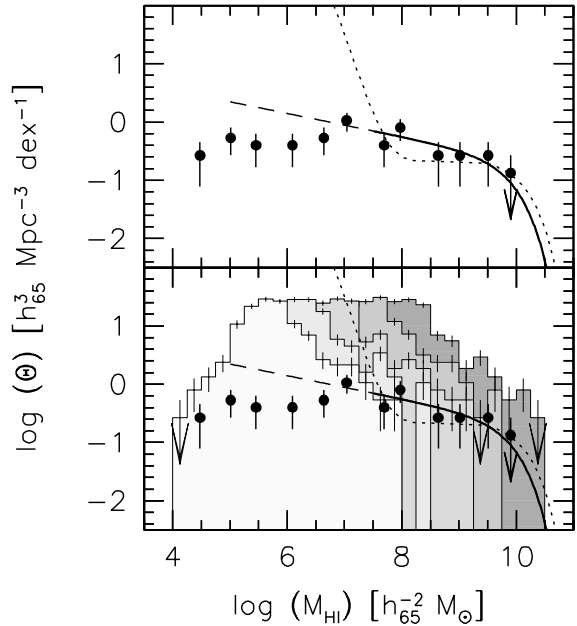} 
\caption{ {\em Top panel:\/} \hi\
mass function of the Local Group (LG). 
 The
points show the space density of LG members containing \hi, after
correcting for incompleteness.  The solid line shows the field \himf\
from Zwaan et al.  (1997), scaled vertically so as to fit the points.
The dotted line is a \himf\ with a steep upturn at the low mass end,
recently proposed by Schneider et al.  (1998)
 {\em Second panel:\/} \hi\ mass functions for extragalactic HVCs.  The
histograms shows the space density of Blitz et al. HVCs if they are put at
the critical radii for gravitational stability assuming different
values of $f$ (from right to left: $f$=0.2, 0.1, 0.05, 0.025, 0.0125).  The
HVC \himf\ is consistent with the field \himf\ if $f\leq 0.02$ and the
median distance $\leq 200$~kpc. }
 \label{himfs.fig}
 \end{figure}

\section{Expected HVC Detections in 21cm Surveys}
 More specific estimates of the expected number of extragalactic HVC
detections can be made by simulating the cloud populations around known
galaxies and groups.  A number of 21cm surveys exist that are capable of
sensing the presence of the extragalactic HVC population in the
outskirts of external groups and galaxies. 

For instance, the Arecibo \hi\ Strip Survey (\ahiss, Sorar 1994, Zwaan
et al.  1997), un unbiased drift-scan survey at two constant declination
strips, is very suitable for assessing the HVC problem.  Zwaan \& Briggs
(2000) use various redshift surveys and catalogs of galaxy groups to
calculate that the strips probe the halos of some 300 galaxies and 14
groups with impact parameters of $\leq 1$~Mpc. The volumes around the
galaxies and groups are filled with synthetic populations of HVCs, with
properties similar to what is proposed by Blitz et al. (1999).  The
group halos are filled with 450 clouds, approximately the number of HVCs
listed by Wakker \& van Woerden (1991), excluding the large complexes
which are excluded from the Blitz et al. analysis. To calculate the
number of clouds around galaxies, the number of clouds associated with
each galaxy is scaled in direct proportion to the ratio of the galaxy
luminosity compared to the integral LG luminosity.

The number of expected detections was calculated taking into account the
column density, size, and velocity width of the HVC population as
measured by Wakker \& van Woerden (1991).  Several radial distribution
functions for the clouds have been tested, but this does not seem to
have a strong influence on the expected number of detections: $\sim 70$
HVC detections in groups and $\sim 250$ detection around galaxies should
have been made.  Note that the group member galaxies have not been removed from
the list of galaxies, so these numbers should not be simply
added to predict a
total number of detections. The analysis is sensitive to clouds with
\hi\ masses $>10^7~\msol$, not only to the most massive ones. 

These predictions are in sharp contrast with the results of the
\ahiss\ analysis: all \hi\ detections away from the Zone of Avoidance
could be optically identified with galaxies containing stars.  Other
surveys have turned up a few instances of dark \hi\ clouds, but these
are all found to be confined to the gravitational potential of a bright
optical galaxy, and are much too scanty to agree with the numbers
calculated above.  Kilborn et al.  (2000) report on the first discovery
of an isolated extragalactic \hi\ cloud without optical identification,
but this cloud too might be explained as a very high velocity cloud
in the outskirts of the Milky Way-LMC system.

\section{QSO Absorption Line Statistics}
 A completely independent way of investigating the idea of extragalactic
HVCs is provided by QSO absorption line statistics.  Charlton, Churchill
\& Rigby (2000) showed that the incidence of Mg~{\sc ii} and Lyman limit
absorption systems is in conflict with the idea that extragalactic
groups contain several hundreds of clouds with typical \hi\ masses of
$10^7\msol$. 
We illustrate this same point here by plotting the column density
distribution function of \hi\ (\fnhi) for the proposed HVC population. 
This function describes the chance of finding an absorber of a certain
\hi\ column density along a random line of sight per unit distance. 

For the HVCs, we again use the Wakker \& van Woerden catalog.  
 The area covered by each cloud is calculated from its measured solid
angle $\Omega$ on the sky, and its distance based on the assumption of
virial equilibrium.  An area function $\Sigma(\nhi)$ that describes the
total area in $\rm Mpc^2$ that is covered by the ensemble of clouds as a
function of \nhi\ is calculated by 
binning the clouds in column density, and adding the areas.
The column density distribution function can be calculated from
 \begin{equation}
 \fnhi = \frac{c}{H_0}\frac{\phi^* \Sigma(\nhi)}{d\nhi},
 \end{equation}
 where $\phi^*$ is the space density of groups.  Charlton et al.  (2000)
derive $\phi^*$ from the CfA redshift survey by dividing the number of
identified groups by the total survey volume and find $\phi^*=3\times
10^{-4}$.  This is a conservative estimate since it does not
incorporate a correction for incompleteness of the survey at higher
redshifts.  Ramella et al.  (1999) compose a catalog of groups from the
ESO Slice Project (ESP) redshift survey.  Within the volume of
$1.9\times 10^5 h_{100}^{-3}~{\rm Mpc}^3$ at the effective depth of
$z=0.16$ they identify 231 groups with at least three members.  The
space density of groups within that volume is therefore at least
$1.2\times 10^{-3} h_{100}^{3}~{\rm Mpc}^{-3}$.  If we limit the
calculation to the volume at the sensitivity peak of the survey at
$z=0.1$ the value of $\phi^*$ rises to $2.0\times 10^{-3}
h_{100}^{3}~{\rm Mpc}^{-3}$, which illustrates the effect of
incompleteness.  As a conservative estimate we adopt
$\phi^*=1.2\times 10^{-3} \rm Mpc^{-3}$ for the calculation of \fnhi. 
The resulting \fnhi\ is shown in Figure~2 as a dashed
histogram.  

The same calculation can be performed for individual galaxies.  In this
case we have to integrate over the luminosity function of galaxies, and
the number of clouds associated with each galaxy is again scaled with
the galaxy luminosity compared to the
integral luminosity of the LG.  For the luminosity function we adopt the
average values from different large galaxy redshift surveys (see
Fukugita, Hogan \& Peebles 1998): $M^*_B=19.5~{\rm mag}$, $\alpha=-1.1$
and $\phi^*=0.020~{\rm Mpc}^{-3}$.  The resulting \fnhi\ is shown as the
solid histogram in Figure~2. 

For comparison, the measured values of \fnhi\ are shown as
points with errorbars.  The leftmost points represent the high \nhi\
part of the Lyman $\alpha$ forest measured by Hu et al.  (1995).  The
HVCs cover the Lyman limit regime, the point here is taken from
Petitjean et al.  (1991).  The forest and Lyman limit points are all
derived from high $z$ ($>2$) data, a low $z$ \fnhi\ is not available in
this regime.  The points for damped Lyman $\alpha$ systems at the
highest \nhi\ part are taken from Rao \& Turnshek (2000) and are for
$z\approx 0.8$.  The straight line is the canonical fit to \fnhi\ at
$z=0$ with a slope $-1.5$. As was noted by Blitz et al. (1999), the {\em
shape} of \fnhi\ for HVCs agrees well with the general observed trend.
However, this plot clearly illustrates that the simulated \fnhi\ for
clouds exceeds the measured one by a large factor. 

This points can be made more qualitatively by calculating \dndz, the
number of absorbers per unit redshift, by integrating over \fnhi.  For
the groups we derive $\dndz=6$ and for clouds around individual galaxies
we find $\dndz=27$.  Extrapolating the results of Stengler-Larrea et al. 
(1995) on Lyman limit systems to $z=0$ yields $\dndz=0.25 \pm 0.15$. 
The conclusions that we can draw from this exercise are similar to those
of Charlton et al.: QSO absorption line statistics are inconsistent with
the hypothesis that galaxies and galaxy groups are surrounded by a
large population of $\mhi=10^7\msol$ gas clouds.

\begin{figure}
\plotone{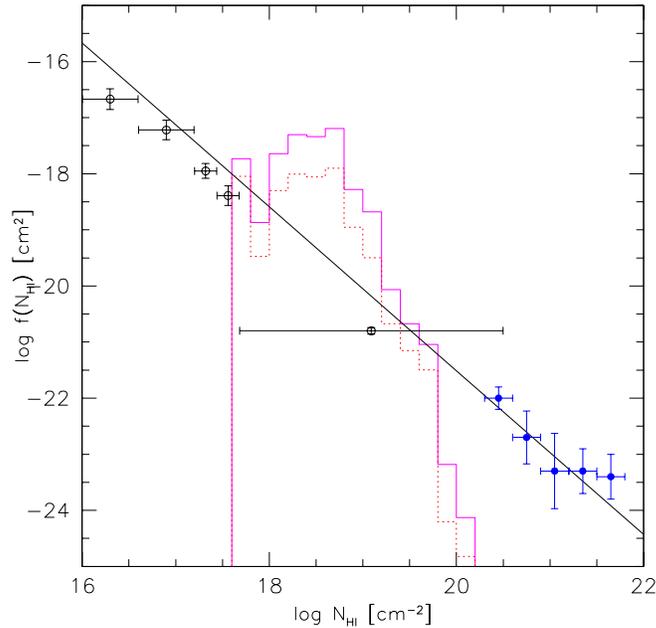}
\caption{The simulated column density distribution function ($\fnhi$) for 
HVCs {\em a)} if they were at typical distances of 1~Mpc distributed
throughout all galaxy groups (dashed histogram); 
{\em b)} if they existed around
all galaxies in the local Universe (solid histogram). The points are
measured values of \fnhi\ from QSO absorption line studies.  
\label{cddf.fig}}
\end{figure}

\section{Discussion}
  The hypothesis that most HVCs are primordial gas clouds with typical \hi\
masses of a few $\times 10^7~\msol$ at distances of $\sim 1$~Mpc from
the Galaxy is not in agreement with observations of nearby galaxies and
groups and with QSO absorption line statistics. 
Blind \hi\ surveys of the extragalactic sky would have detected
these clouds if they exist around all galaxies or galaxy groups in
numbers equal to those suggested for the Local Group.  These results are
highly significant: the Arecibo \hi\ strip survey would have detected
approximately 250 clouds around individual galaxies and 70 in galaxy
groups. Furthermore, the measured incidence of QSO absorption line systems
\dndz\ is at least a factor 20 lower than what would be expected if all
groups and galaxies were surrounded by a collection of \hi\ clouds
similar to what is proposed for the Local Group.

Several additional observations could help identifying the true nature
of HVCs.  Measuring the H$\alpha$ flux in HVCs can yield accurate
distances, provided that the UV ionizing field around the Milky Way is
accurately known (Bland-Hawthorn et al.  1998, Bland-Hawthorn \& Maloney
1999).  Also, determining metal abundances in HVCs is a useful method to
distinguish between different HVC scenarios.  If the clouds are
primordial remnants of the formation of the Local Group they would not
have been contaminated with these heavier elements and would retain a
composition closer to the pristine environment of the early Universe. 
                                                                        
Results to date have been confusing, since different teams report
different measured metallicities, even in the same cloud.  HST
spectra of Complex C taken in the direction of
background source Mrk~290 yield a sulphur abundance ten times
lower than that of the gas layer of the Milky Way (Wakker et al.  2000). 
On the other hand, very recent data from FUSE, probing a different
position in the same cloud, show an abundance of iron approximately half
that found in the solar neighborhood (Murphy et al.  2000). 
Furthermore, Sembach et al.  (2000) report on FUSE observations that
show the first detection in a HVC of O~{\sc vi}, a high ionization
level probably
produced in interactions between the HVC and the halo gas of the Galaxy. 

The conclusion that can be drawn at this moment is that HVCs are
probably a mixture of several species with different origin.  Some are
clearly the result of gravitational interactions, some might be produced
in a Galactic fountain where hot gas is blown out of the Galactic disk,
into the halo, where it cools and precipitates on the Galaxy.  Some
might be primordial gas clouds distributed throughout the Local Group. 
Braun \& Burton (1999, 2000) limit their calculations to a subsample of
some 80 compact HVCs, and place them at a median distance of 650 kpc at
which the \hi\ mass would be $\sim 10^7\msol$.  For such a population,
the limits from both QSO absorption line results and 21cm surveys are
much weaker.  Future deep 21cm surveys that reach minimal detectable
\hi\ masses of $10^6~\msol$ in nearby groups will provide the definitive
answer.


\begin{references}
\reference Bland-Hawthorn, J., \& Maloney, P.~R. 1999, \apj, 510, L33
\reference Bland-Hawthorn, J., Veilleux, S., Cecil, G.~N., Putman, M.~E., 
	Gibson, B.~K., \& Maloney, P.~R. 1998, \mnras, 299, 611
\reference Blitz, L., Spergel, D.~N., Teuben, P.~J., Hartmann, D., 
	\& Burton, W.~B. 1999, \apj, 514, 818   
\reference Blitz, L. \& Robishaw, T. 2000,  astro-ph/0001142   
\reference Braun, R., \& Burton, W.~B. 1999, \aap, 341, 437
\reference Braun, R., \& Burton, W.~B.  2000, astro-ph/0004033
\reference Charlton, J.~C., Churchill, C.~W., \& Rigby, J.~R. 2000,
	astro-ph/0002001
\reference Fukugita, M., Hogan, C.~J., \& Peebles, P.~J.~E. 1998, \apj,
	503, 518
\reference Garcia, A.~M. 1993, A\&AS, 100, 47    
\reference Hu, E.~M., Kim, T., Cowie,  L.~L., Songaila, A. \& Rauch, 
	M. 1995, \aj, 110, 1526 
\reference Hulsbosch, A.~N.~M. 1975, \aap, 40, 1         
\reference Kilborn, V. et al. 2000, astro-ph/0005267
\reference Klypin, A.~A., Kravtsov, A.~V., Valenzuela, O., \& 
	Prada, F. 1999, \apj, 522, 82   
\reference Mateo, M.~L. 1998, \araa, 36, 435      
\reference Moore, B., Ghigna, F., Governato, F., Lake, G., Stadel J., 
	Tozzi, P. 1999, \apj, 524, L19
\reference Murphy, E. et al. 2000, astro-ph/0005408
\reference Oort, J.~H., Bull. Astr. Inst. Neth. 1966, 18, 421
\reference Petitjean, P., Webb, J.~K., Rauch, M., Carswell, R.~F. \& 
	Lanzetta, K. 1993, \mnras, 262, 499 
\reference Putman, M.~E.  et al. 1998, Nature, 394, 752  
\reference Rao, S.~M., Turnshek, D.~A. 2000, astro-ph/9909164
\reference Ramella, M. et al. 1999, \aap, 342, 1
\reference Schneider, S.~E., Spitzak, \& J.~G., Rosenberg, J.~L. 
	1998, \apj, 507, L9   
\reference Sembach, K.~R. et al. 2000, astro-ph/0005012
\reference Sorar, E.  1994, Ph.D.  Thesis, University of Pittsburgh            
\reference Spitzak, J.~G., \& Schneider, S.~S. 1999, \apjs, 119, 159      
\reference Stengler-Larrea, E.~A., et al. 1995, \apj, 444, 64
\reference Wakker, B.~P., \& van Woerden,  H. 1991, \aap, 250, 509 
\reference Wakker, B.~P., \& van Woerden, H. 1997, \araa, 35, 217        
\reference Wakker, B.~P. et al. 1999, Nature, 402, 388 
\reference van Woerden, H., Schwarz, U.~J., Peletier, R.~F., Wakker, 
	B.~P., \& Kalberla, P.~M.~W. 1999, Nature, 400, L138
\reference Zwaan, M.~A., Briggs, F.~H., Sprayberry, D., \& 
	Sorar, E. 1997, \apj, 490, 173  
\reference Zwaan, M.~A., \& Briggs, F.~H. 2000, \apj, 530, L61
\end{references}
\end{document}